# Negative Poisson's Ratio in 1T-Type Crystalline Two-Dimensional Transition Metal Dichalcogenides

Liping Yu*, Qimin Yan, and Adrienn Ruzsinszky

Department of Physics, Temple University, Philadelphia, Pennsylvania 19122, USA

## Abstract

Materials with a negative Poisson's ratio, also known as auxetic materials, exhibit unusual and counterintuitive mechanical behavior – becoming fatter in cross-section when stretched. Such behavior is mostly attributed to some special re-entrant or hinged geometric structures regardless the chemical composition and electronic structure of a material. Here, using first principles calculations, we report a new class of auxetic single-layer two-dimensional (2D) materials, i.e., the 1T-type monolayer crystals of groups 6-7 transition-metal dichalcogenides, $MX_2$ (M = Mo, W, Tc, Re; X = S, Se, Te). These materials have a crystal structure distinct from all other known auxetic materials. They exhibit an intrinsic in-plane negative Poisson's ratio, which is dominated by the electronic effects. We attribute the occurrence of such auxetic behavior to the strong coupling between the chalcogen $p$ orbitals and the intermetal $t_{2g}$-bonding orbitals within the basic triangular pyramid structure unit. The unusual auxetic behavior in combination with other remarkable properties of monolayer 2D materials could lead to novel multi-functionalities.

---

* Corresponding author: L.Y. (email: yuliping@gmail.com )

## Introduction

The Poisson's ratio of a material characterizes its response to uniaxial load and is given by $\nu_{ab} = -\varepsilon_b/\varepsilon_a$, where $\varepsilon_a$ is an applied strain in the $a$-axis direction and $\varepsilon_b$ is the resulting strain in a transverse $b$-axis direction. Counter-intuitively, negative Poisson's ratio (auxetic) materials[1], expand laterally when stretched and contract laterally when compressed. They can lead to enhanced mechanical properties such as shear modulus[2], indentation resistance[3], and fracture toughness[4]. The unusual auxetic effect itself and concomitant enhancements in other material properties offers enormous potential in many technologically important applications[5-7] such as biomedicine[8], sensors[9], fasteners[10], and protective equipments[11].

Auxetic effect has been reported in a number of natural and man-made materials and structures in bulk form[5,6,12,13], for example, cubic metals[14,15], α-cristobalite ($SiO_2$)[16], α-$TeO_2$[17], the zeolite mineral natrolite[18], honeycombs[19], foams[7], microporous polymers[20,21], composites[22,23], ceramics[24], molecular auxtics[25], metal-organic frameworks[26], bucklicrystals[27], and origami structures[28-30]. Geometric considerations dominate the literature in understanding such auxetic effects and designing new auxetic materials. For most of these auxetic materials, the auxetic effect is explained by some special re-entrant structure or the crystal structure that can be viewed as being made up of rigid building blocks linked by flexible hinges[1,19,31-33], independent of their chemical composition and electronic structure.

Auxetic effect has also been recently reported in several monolayer 2D materials. For example, the out-of-plane negative Poisson's ratio was discovered in phosphorene[34,35], GeS[36], and monolayer arsenic[37]. The in-plane negative Poisson's ratio was also predicted in borophene[38] and three theoretically proposed but not-yet-synthesized materials (i.e., the penta-graphene[39,40], $h\alpha$-silica[41], and $Be_5C_2$[42]). Similar to that in the bulk auxetic materials, the auxetic behavior in these 2D materials is also considered to originate mainly from the puckered or buckled crystal structure.

In this study, using quantum mechanical first-principles calculations (see methods), we report a new class of auxetic single-layer 2D materials with an intrinsic in-plane negative Poisson's ratio. They differ from other known auxetic materials not only in their crystal structure but also in the microscopic origin of auxetic behavior. These materials are the 1T-type crystalline monolayers of groups 6-7 transition metal dichalcogenides, 1T-$MX_2$ (M = Mo, W, Tc, Re; X = S, Se, Te). In contrast to those known bulk or 2D auxetic materials, the in-plane auxetic behavior discovered in groups 6-7 1T-$MX_2$ cannot be explained merely from their geometric structure because the non-auxetic behavior is also found in other groups of $MX_2$ compounds with the same 1T-type structure. This dichotomy between auxetic and non-auxetic behavior in the 1T-$MX_2$ compounds is explained by their distinct electron structures. The in-plane stiffness of those 1T-$MX_2$ materials is predicted to be order of $10^2$ GPa, at least three orders of magnitude higher than man-made auxetic materials. The high in-plane stiffness and the auxetic behavior in combination with other remarkable electronic and optoelectronic properties of the single-layer 2D materials[43] could lead to novel multi-functionalities such as nanocale auxetic electrodes and sensors.

## Results

**Crystal structure**. The single layers of 2D transition metal dichalcogenides are formed by a hexagonally packed layer of metal (M) atoms sandwiched between two layers of chalcogen (X) atoms (Figure 1). Each chalcogen atom forms the apex of a triangular pyramid that has three metal atoms at its base. The symmetry of the chalcogen array about each metal atom is either octahedral or trigonal prismatic. The former is often referred to as the 1T phase whereas the latter as the 1H phase. Depending on the combination of the metal and chalcogen elements, one of the two phases is thermodynamically preferred. Most group-6 $MX_2$ compounds thermodynamically prefer the 1H phase[44], but the metastable 1T phase is also observed[45-47]. For other groups of layered $MX_2$ compounds, most crystallize in the high-symmetry 1T or low-symmetry distorted-1T phase[44,48]. The 1H-$MX_2$ compounds are known to be non-auxetic in the plane due to their hexagonal in-plane crystalline structure. We hence focus on 42 monolayer $MX_2$ compounds in the high-symmetry 1T-phase (Table 1).

**Poisson's Ratio results.** Figure 2a shows our calculated Poisson's ratio results ($v_{ab}$) for 42 1T-$MX_2$ compounds in the *b*-axis direction subjected to a 5% tensile strain applied along the *a*-axis direction. Remarkably, we find that the sign of Poisson's ratio strongly depends on the d-electron count. All twelve 1T-$MX_2$ compounds from group 6 ($d^2$) and group 7 ($d^3$) exhibit negative Poisson's ratios, ranging from −0.03 to −0.37. Seven of them (i.e., $TcTe_2$, $ReTe_2$, $WTe_2$, $WSe_2$, $MoSe_2$, $ReS_2$, and $TcS_2$) have a Poisson's ratio less than −0.1, higher in magnitude than that of borophene (-0.04 along *a* and -0.02 along *b*)[38], rendering them more promising candidates for specific applications in mechanical nanodevices. For other groups of 1T-$MX_2$ compounds, we find positive Poisson's ratios ranging from 0.09 to 0.53.

Fig.2bc shows our calculated Poisson's ratios ($v_{ab}$ and $v_{ba}$) as a function of applied strain in two example compounds, non-auxetic $ZrS_2$ and auxetic $MoS_2$. For both compounds, Poisson's ratio varies slowly as applied strain goes from -5% to 5%, suggesting a dominant linear elastic behavior within the strain range considered. (Note the Poisson's ratio at a large strain (i.e., > 5% or < -5%) may strongly depend on the strain. This behavior is not pursued in this work since such large strains are often experimentally inaccessible.) The small differences between $v_{ab}$ and $v_{ba}$ reflect a nearly isotropic auxetic or non-auxetic behavior inside the 1T-structure plane. Therefore, the d-electron count dependence of the sign of Poisson's ratio as shown in Fig. 2a does not change with respect to the amount of the applied strain within the linear elastic range and the loading direction inside the plane.

**Stiffness.** To compare the stiffness (Young's modulus) of a single-layer material with bulk materials, we calculate its 3D in-plane stiffness ($Y_{3D}$) from 2D in-plane stiffness ($Y_{2D}$) and effective layer thickness ($t$) via $Y_{3D} = Y_{2D}/t$. The $Y_{2D}$ is directly derived from first-principles total energies as a function of uniaxial strain. The effective layer thickness $t$ can also be uniquely determined from first-principles calculated bending energy[49]. Here for simplicity we approximate $t$ as $t = t_0 + 0.8$ Å, where $t_0$ is the distance between the top and bottom chalcogen atom layers and the 0.8 Å is the total effective decay length (0.4 Å in each layer side) of electron density into the vacuum. The 0.8 Å is derived from the first-principles calculated layer thickness for 1H-$MoS_2$[49]. $MSe_2$ and $MTe_2$ may have different decay lengths than $MS_2$. However, such difference should be less than one time of magnitude. Hence, using a different decay length does not induce one time of magnitude difference in the calculated 3D in-plane stiffness.

Table 1 shows that the 3D in-plane stiffness of almost all 1T-$MX_2$ compounds lies in between 100 and 300 GPa. Among the auxetic $d^2$-$d^3$ 1T-$MX_2$ compounds, $WS_2$ and $ReSe_2$ are stiffest, having a stiffness of ~290 GPa; $TcTe_2$ is softest, having a stiffness of ~80 GPa. Man-made auxetic materials typically have a stiffness in the range from ~$10^{-5}$ GPa to ~1 GPa, and naturally occurring auxetic bulk solids exhibit a stiffness of ~$10^1$-$10^2$ Pa[50]. Therefore, even considering the uncertainty of our calculated 3D stiffness (less than one order of magnitude) that may be caused by using different approximations for effective layer thickness, the 3D stiffness values predicted for 1T-$MX_2$ compounds are among the highest in the naturally occurring crystalline solids and are at least three orders of magnitude higher than man-made auxetic materials.

The fact that both auxetic and non-auxetic materials are found in the same 1T-structure type implies that the auxetic effect is not a purely geometric property. The d-electron count dependence of electronic structure must be involved. In the 1T-structure, the $d$ orbitals of the octahedrally coordinated transition metal split into two groups, $d_{xy,yz,zx}$ ($t_{2g}$) and $d_{x2-y2,z2}$ ($e_g$). In what follows, we shall show that (i) transition metals interact with each other through $t_{2g}$-orbital coupling, and (ii) the coupled $t_{2g}$ orbitals are further coupled with the "lone-pair" electrons of chalcogen atoms. It is the gradual filling of such $t_{2g}$-p hybridized bands that leads to the different behavior of Poisson's ratio.

**Intermetal $t_{2g}$-orbital coupling.** In the ideal 1T phase, the M-centered octahedra share edges, forming three one-dimensional M-chains along the directions of lines $y = x$, $y = z$, and $z = -x$, respectively, within the local reference frame of the octahedra (Fig. 1b). The metal atoms can interact with each other through the coupling between their $t_{2g}$ orbitals. This coupling gives rise to $t_{2g}$-bonding states and $t_{2g}$-antibonding states, with no energy gap in between due to the weak coupling nature. The $t_{2g}$-states are mostly located within the gap between the bonding and antibonding bands of the M-X bonds (Fig.1c).

The progressive filling of these $t_{2g}$ bands from group 4 ($d^0$) to group 10 ($d^6$) species leads to different M-M bonding or antibonding character at the Fermi level. In $d^1$-$d^3$ 1T-MX$_2$, the Fermi level crosses the $t_{2g}$-bonding states; the highest occupied bands close to the Fermi-level thus exhibit a stronger bonding character as we go from $d^1$ to $d^3$. This bonding character attracts the metal atoms towards each other, leading to an intermetal distance shorter than that in the ideal 1T structure. In $d^5$-$d^6$ 1T-MX$_2$, since the $t_{2g}$ bonding states can accommodate up to six electrons (three from each metal), all $t_{2g}$-bonding states are filled and the Fermi level crosses the $t_{2g}$-antibonding states. Hence the highest occupied bands in the vicinity of the Fermi-level exhibit anti-bonding character, repelling metal atoms from each other.

The existence of the intermetal $t_{2g}$-orbital interactions is reflected by the d-electron count dependence of the M-X-M bond angles (∠MXM) as illustrated in Fig.1d. The ideal 1T phase has regular octahedra with ∠MXM = 90º. In the $d^0$ 1T-MX$_2$ compounds, the ∠MXM deviates least from 90º. This is expected since all $t_{2g}$ states are almost completely unoccupied and the intermetal d-d interaction is marginal. For the $d^1$-$d^3$ 1T-MX$_2$, all have acute ∠MXM, decreasing with the increasing d-electron count. This trend arises from the increasing intermetal $t_{2g}$-bonding character in going from $d^1$ to $d^3$, which shortens the intermetal distance. In the $d^5$-$d^6$ 1T-MX$_2$, the ∠MXM jumps up to over 90º, consistent with the intermetal $t_{2g}$-antibonding character.

Fig.1d also shows that the chalcogen atoms have minor effect on ∠MXM compared with the transition metals with different d-electron counts, but a trend can still be observed: the ∠MXM decreases with increasing atomic number of the chalcogen. For example, the ∠MXM of TiS$_2$, TiSe$_2$, and TiTe$_2$ decreases from 89.6º to 88.2º to 86.0º. This trend is not associated to the intermetal $t_{2g}$-orbital interaction; instead it is intrinsic to the spatial distribution of the lone-pair charge density relative to that of the M-X bonds around the chalcogen.

**$t_{2g}$-p orbital coupling.** The intermetal $t_{2g}$ orbitals are further coupled with chalcogen p orbitals in 1T-MX$_2$. It can be seen from their projected density of states (DOS) as shown in Figure 3. In the 4d transition metal disulfides with the ideal 1T-structure, we find that the DOS of sulfur 3p and metal $t_{2g}$ states overlap, as manifested by their similar DOS peak shapes and positions in energy. The $t_{2g}$-p orbital overlap is marginal in $d^0$ ZrS$_2$, but it increases quickly in going from $d^1$ NbS$_2$ to $d^6$ PdS$_2$. This trend is clear not only in the energy range from -12 eV to -7 eV, where the major peaks of 3p-DOS are located, but also near the Fermi level.

The $t_{2g}$-p orbital interaction is attractive because the X ligand has one lone electron pair and acts as a sigma donor. In $d^1$-$d^3$ MX$_2$, the $t_{2g}$-p coupling force draws atom X towards the intermetal bond centers, because the $t_{2g}$ states are the intermetal bonding states spreading over the M-M bond centers. In $d^5$-$d^6$ MX$_2$, the $t_{2g}$-p coupling force attracts atoms M and X towards each other, because the $t_{2g}$ states are antibonding and localized near the metal atoms. The d-electron count dependence of $t_{2g}$-p interaction direction plays a key role in determining the structure deformation presented below.

**Deformation mechanism.** To understand the microscopic origin of Poisson's ratios, let us now look into the resulting structural relaxation subjected to a tensile strain applied along the *a*-axis. Due to the centrosymmetric nature of the 1T phase, the whole relaxation process manifests itself in the triangular pyramid unit as illustrated in Figure 4. For the stretch along the $M_1$-$M_3$ axis (i.e., axis *a*), the resulting relaxation involves only atoms $M_2$ and X moving inside the Q-X-$M_2$ plane. Hence two relations always hold during relaxation: $d_{M_1M_2} = d_{M_2M_3}$ and $\angle M_1XM_2 = \angle M_3XM_2$.

We analyze the relaxation process by decomposing it into three consecutive steps: (i) atom X relaxes along the line Q-X, (ii) atom X rotates around the $M_1$-$M_3$ axis, and (iii) atom $M_2$ relaxes along the line Q-$M_2$. In the first two steps, the lattice constant *b* is fixed to the value found in the relaxed strain-free 1T-structure. In the third step, the *b* varies as atom $M_2$ moves along the Q-$M_2$ line, leading to different Poisson's ratio behavior.

Fig.4 shows the detailed structural relaxation in the three consecutive steps described above for 1T-$MX_2$ with $\angle QXM_2 < 90º$ and with $\angle QXM_2 > 90º$ separately. Each step can be understood in the way that atom X (or atoms X and $M_2$) relaxes to conserve the M-X bond length ($d_{MX}$) since $d_{MX}$ is energetically dominant. After the first two steps of the relaxation, it can be seen that (i) both $d_{M_1M_2}$ (also $d_{M_2M_3}$) and $\angle M_1XM_2$ and $\angle M_3XM_2$ (supplementary Figure 1) increase in all 1T-$MX_2$ compounds no matter whether $\angle QXM_2$ is larger or smaller than 90º, and (ii) $\angle XQM_2$ increases in the 1T-$MX_2$ with $\angle QXM_2 < 90º$ but decreases in the 1T-$MX_2$ with $\angle QXM_2 > 90º$ (Supplementary Figure 1). The changes in $d_{M_1M_2}$ and $d_{M_2M_3}$, $\angle M_1XM_2$ and $\angle M_2XM_3$, and $\angle XQM_2$, thus store the strain energy, which will be partially released in the subsequent third step relaxation.

The third step relaxation determines the sign of Poisson's ratio. The negative Poisson's ratio of $d^2$-$d^3$ $MX_2$ can be attributed to the strong $t_{2g}$-p orbital coupling. Such strong coupling implies a large amount of strain energy stored in the decreased $\angle XQM_2$ after the second step. This part of strain energy will be released in this third step through atom $M_2$ relaxing along the increased *b*-lattice direction, leading to a negative Poisson's ratio. The strength of $t_{2g}$-p orbital coupling depends not only the d-electron count of the transition metal but also on the chalcogen atom. This dependence explains why the Poisson's ratio of the compounds from same $d^2$ or $d^3$ group also differs from one another as shown in Fig.2a.

For $d^0$-$d^1$ $MX_2$, the positive Poisson's ratio results from the marginal or weak intermetal $t_{2g}$-coupling and $t_{2g}$-p coupling. Such weak couplings imply that the strain energy stored in $d_{M_1M_2}$ and $d_{M_2M_3}$ and $\angle XQM_2$ is also marginal or small. The major strain energy that can be released in the third step is thus stored in the increased $\angle M_1XM_2$ and $\angle M_2XM_3$. Therefore, it is energetically favorable that atom $M_2$ relaxes in the *b*-decreasing direction, reducing the increase in ∠M$_1$XM$_2$ and ∠M$_2$XM$_3$, and resulting in a positive Poisson's ratio. For d$^5$-d$^6$ MX$_2$, the positive Poisson's ratio originates from the fact that the t$_{2g}$-p coupling aligns along the M-X bond and does not energetically affect the change in ∠XQM$_2$. In other words, the strain energy stored in the decreased ∠XQM$_2$ is small. Since the t$_{2g}$-antibonding is also generally weak, the relaxation of atom M$_2$ is energetically favorable in the *b*-decreasing direction, giving rise to a positive Poisson's ratio. This deformation mechanism is similar to that in d$^0$-d$^1$ compounds.

Simply saying, the negative Poisson's ratio in d$^2$-d$^3$ MX$_2$ results from the strong attractive coupling between the intermetal t$_{2g}$-bonding states and the X p states, which prevents atoms X and M$_2$ relaxing toward the ∠XQM$_2$-increasing direction. The positive Poisson's ratio arises from lack of such strong t$_{2g}$-p coupling in other groups of 1T-MX$_2$.

## Discussion

The monolayer MX$_2$ materials involve transition metals where strong correlation effects may be not well captured by the new SCAN meta-GGA functional. To check the robustness of our results, we also calculated the Poisson's ratio for 12 d$^2$-d$^3$ MX$_2$ by using the HSE06 hybrid functional[51]. The results are summarized in supplementary Table 1. It shows that the Poisson's ratio of eight 1T-MX$_2$ compounds (i.e., MoSe$_2$, MoTe$_2$, WSe$_2$, WTe$_2$, TcTe$_2$, ReS$_2$, ReSe$_2$, ReTe$_2$) remains negative, whereas for other four compounds (i.e., MoS$_2$, WS$_2$, TcS$_2$, TcSe$_2$) their Poisson's ratio changes the sign from negative to slightly positive, which is still very interesting and useful for applications. Although it is found that the SCAN lattice constants agree better with experiment than the HSE06 ones for most of the compounds listed in this table, it is uncertain whether SCAN predicts a more accurate Poisson's ratio than HSE06 since the semilocal SCAN functional could make larger density-driven error in the energy than HSE06 does for the system under stretching[52]. This uncertainty calls for experimental validation and further theoretical study. Nevertheless, the auxetic behavior we find is robust in most of the d$^2$-d$^3$ MX$_2$ compounds. The less negative Poisson's ratio predicted by HSE06 (supplementary Table 1) further indicates that the auxetic behavior originates from the strong p-d coupling. In general, compared with the semi-local SCAN functional, HSE06 yields more localized metal d and chalcogen p orbitals and hence the weaker hybridization between them, which leads to less negative Poisson's ratios in HSE06.

Our predicted in-plane auxetic behavior is intrinsic in the 1T-structure without any external engineering and occurs in the elastic region. This is different from the extrinsic auxetic behavior reported in the epitaxial oxide thin-film[53,54] and the engineered 2D materials such as the wrinkled graphene[55], graphane[56], and borophane[57]. Recently, the "negative Poisson's ratio" was also reported in metal nanoplates[58], pristine graphene[59], and semi-fluorinated graphene[60]. The "negative Poisson's ratios" claimed there actually correspond to the ratio calculated from $\nu_{ab} = -\partial\varepsilon_a/\partial\varepsilon_b$, not from $\nu_{ab} = -\varepsilon_a/\varepsilon_b$ (the original definition of Poisson's ratio). The latter tells whether the material is auxetic (i.e. becoming fatter when stretched), whereas the former does not. Negative of the former does not imply negativity of the latter (auxetic behavior). Indeed, these three materials exhibit non-auxetic behavior defined by $\nu_{ab} = -\varepsilon_a/\varepsilon_b$ at least in the elastic region, hence differing from our discovered auxetic 2D materials.

Finally, it is noteworthy that the auxetic behavior of $d^2$-$d^3$ $MX_2$ compounds is predicted in the high-symmetry 1T-phase. This phase is known to be metastable or dynamically unstable in both $d^2$ and $d^3$ $MX_2$ compounds[44,61-63]. However, experimentally, relevant phase diagrams of monolayer materials differ from those of bulk materials. The kinetic barriers between the different phases of monolayers may arise and be affected by many external factors such as interfaces, underlying substrate, temperature, strain, and impurities. Therefore, it is not uncommon to observe the undistorted 1T-phase synthesized experimentally. For instance, although no kinetic barrier is found from first-principles calculations between the unstable 1T phase and dynamically stable distorted-1T phase, the undistorted 1T monolayer structures of $MoS_2$, $MoSe_2$, $WS_2$, and $WSe_2$ are observed from the exfoliation using Li-intercalation method[45,64]. For $MoS_2$, the coexistence of 1T and 1H domains is also observed in the same monolayer[46,47]. Such heterogeneous monolayers with auxetic and non-auxetic domains are particularly intriguing since they could lead to novel functionality.

## Methods

All calculations were performed using density functional theory (DFT) and the plane-wave projector augmented-wave (PAW)[65] method as implemented in the VASP code[66]. The new SCAN (strongly constrained and appropriately normed) meta-generalized gradient approximation was used[67,68]. SCAN is almost as computationally efficient as PBE-GGA functional, yet it often matches or exceeds the accuracy of the more computationally expensive hybrid functionals in predicting the geometries and energies of diversely bonded systems[68]. Supplementary Table 2 shows our calculated lattice constants for 1T-$MX_2$ compounds. They agree very well with available experimental data[44], especially for groups 4-7 1T-$MX_2$ whose errors are within 1%. An energy cutoff of 500 eV was used. The monolayer structure is modeled in an orthorhombic supercell that contains two formula units (Fig.1a) and a 20Å vacuum space inserted in the out-of-plane direction. A 24×14×1 k-point grid was used to sample the Brillouin zone during structure relaxation. All atoms were fully relaxed until their atomic forces were less than 0.005 eV/Å. The effects of spin-orbit coupling on the structural deformation are considered to be minor and hence not included in our study.

The Poisson's ratio is calculated from the engineering strain ($\varepsilon$), which is defined as the change in length $\Delta L$ per unit of the original length $L$, i.e., $\varepsilon = \Delta L/L$. The applied uniaxial strain is realized in our calculations by fixing the lattice parameter to a value different from its equilibrium value during structural relaxation. The resulting strain in the transverse direction is extracted from the fully relaxed structure subjected to an applied strain.

## Acknowledgements

We thank John P. Perdew for valuable scientific discussions and comments on the manuscript. The authors also thank Richard C. Remsing and Jefferson E. Bates for the comments on the manuscript. This research was supported as part of the Center for the Computational Design of Functional Layered Materials (CCDM), an Energy Frontier Research Center funded by the U.S. Department of Energy (DOE), Office of Science, Basic Energy Sciences (BES), under Award #DE-SC0012575. This research used resources of the National Energy Research Scientific Computing Center, a DOE Office of Science User Facility supported by the Office of Science of the U.S. Department of Energy under Contract No. DE-AC02-05CH11231. This research was also supported in part by the National Science Foundation through major research instrumentation grant number CNS-09-58854.

## Contributions

L.Y. designed the project, performed the calculations and wrote the manuscript. Q.Y. and A.R. contributed to analyzing the results and writing the manuscript.

## Competing financial interests

The authors declare no competing financial interests.

**Table 1.** Predicted in-plane stiffness (Young's Modulus) for 42 monolayer 1T-MX$_2$ compounds. The 3D in-plane stiffness ($Y_{3D}$) is calculated from 2D in-plane stiffness ($Y_{2D}$) divided by the effective layer thickness $t$ given by $t = t_o + 0.8$ Å, where $t_o$ is the distance between the top and bottom chalcogen atom layers.

| $M^{4+}$ | $X_2$ | 3D in-plane stiffness $Y_{3D}$ (GPa) | | | 2D in-plane stiffness $Y_{2D}$ (Nm$^{-1}$) | | |
|---|---|---|---|---|---|---|---|
| | | -S$_2$ | -Se$_2$ | -Te$_2$ | -S$_2$ | -Se$_2$ | -Te$_2$ |
| d$^0$ | Ti | 236 | 183 | 108 | 85 | 70 | 46 |
| | Zr | 210 | 182 | 103 | 77 | 71 | 44 |
| | Hf | 233 | 200 | 117 | 85 | 77 | 50 |
| d$^1$ | V | 263 | 226 | 159 | 97 | 88 | 67 |
| | Nb | 225 | 180 | 127 | 87 | 73 | 56 |
| | Ta | 265 | 215 | 129 | 101 | 85 | 57 |
| d$^2$ | Mo | **261** | **249** | **205** | 103 | 104 | 92 |
| | W | **289** | **225** | **199** | 113 | 94 | 88 |
| d$^3$ | Tc | **232** | **244** | **77** | 94 | 104 | 34 |
| | Re | **222** | **286** | **157** | 90 | 123 | 71 |
| d$^5$ | Co | Non-Layered Structure | | 164 | Non-Layered Structure | | 59 |
| | Rh | | | 98 | | | 37 |
| | Ir | | | 121 | | | 45 |
| d$^6$ | Ni | 287 | 248 | 126 | 89 | 80 | 44 |
| | Pd | 232 | 193 | 180 | 77 | 66 | 63 |
| | Pt | 296 | 244 | 208 | 96 | 82 | 74 |

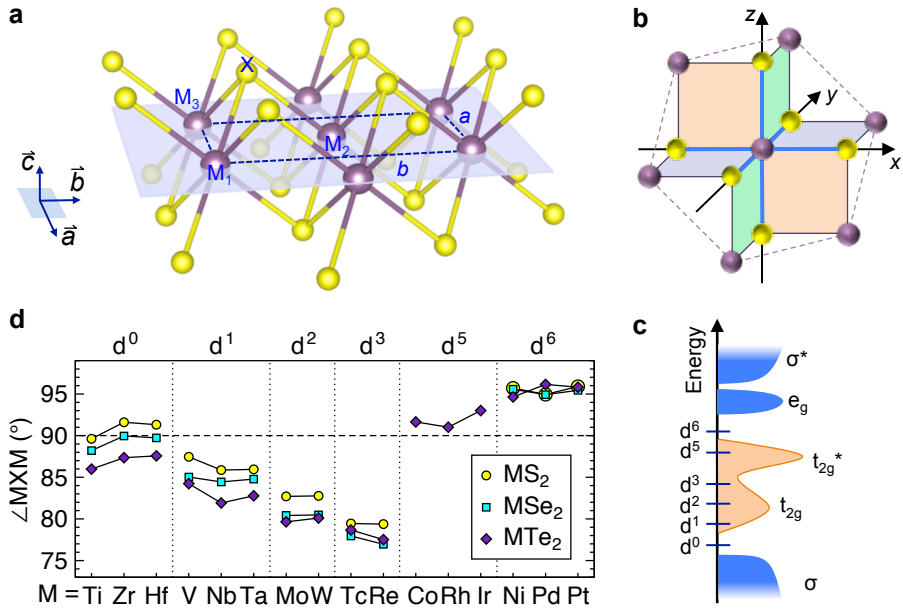

**Figure 1 | Structure of monolayer 1T-MX$_2$. a,** Crystal structure. The basic X-M$_1$-M$_2$-M$_3$ triangular pyramid unit is marked. The rectangular outline displays the unit cell adopted in our calculation. It contains two MX$_2$ formula units. **b,** Local structure of M-centered octahedron. The metal atoms form three one-dimensional chains in the directions of y = x, z=y, and z=-x in the local reference frame. The M-M interaction is through the t$_{2g}$-orbital coupling. **c**, Schematic configuration of density of states showing the gradual filling of d orbitals from group 4 (d$^0$) to group 10 (d$^6$) 1T-MX$_2$. The horizontal bars denote the corresponding Femi level of the system. t$_{2g}$ and t$_{2g}$* correspond to the intermetal t$_{2g}$-bonding and t$_{2g}$-antibonding states, respectively. **d,** Predicted M-X-M bond angles in the relaxed structure of strain-free 1T MX$_2$. Note, in the triangular pyramid as shown Fig.1a, ∠M$_1$XM$_2$ = ∠M$_2$XM$_3$ = ∠M$_3$XM$_1$ = ∠MXM.

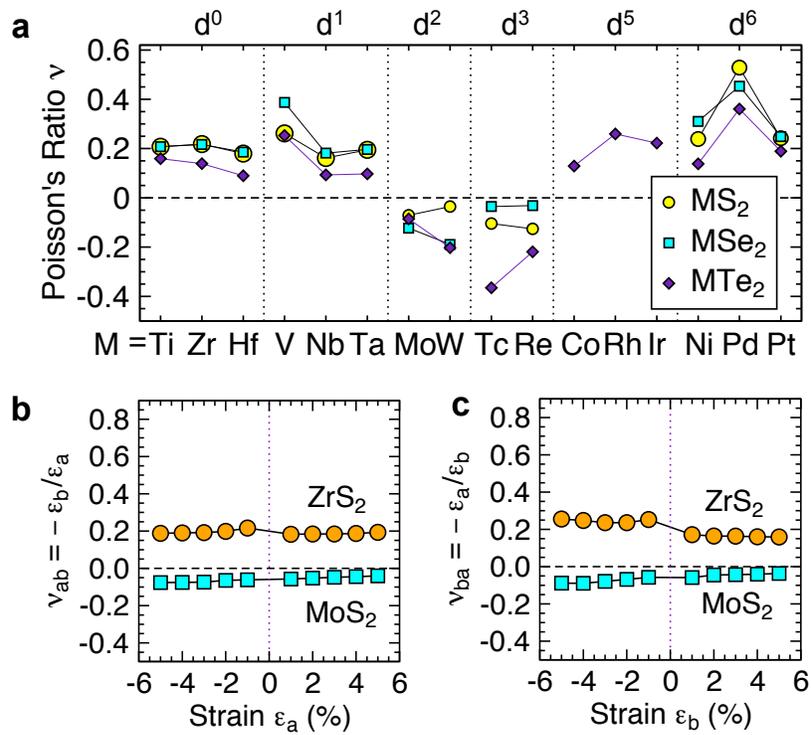

**Figure 2 | Poisson's ratios. a**, Poisson's ratio, $v_{ab} = -\varepsilon_b/\varepsilon_a$, calculated for a 5% strain applied along the a-axis (i.e., $\varepsilon_a$ = 5%). **b-c**, Poisson's ratios for $ZrS_2$ and $MoS_2$ as a function of strain applied along the *a*-axis (b) and *b*-axis (c).

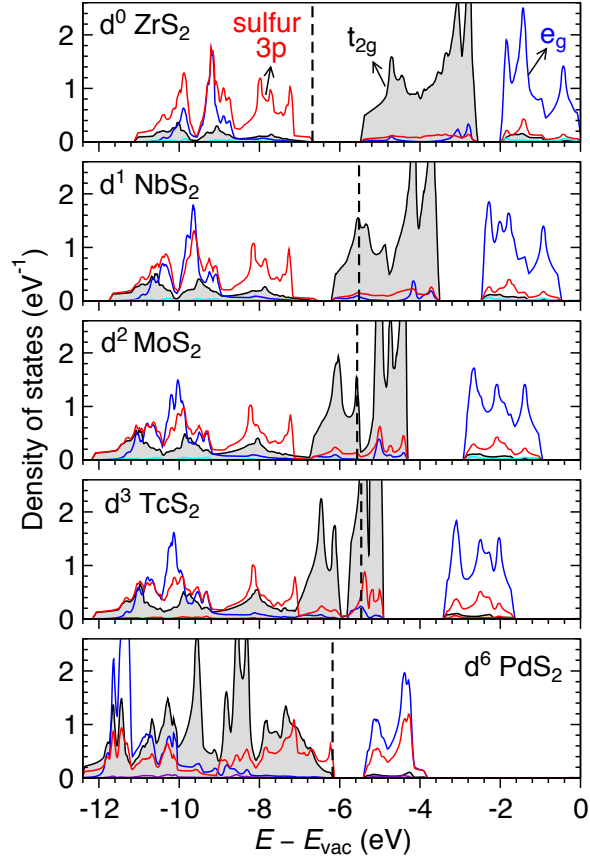

**Figure 3 | Density of states of 4d MS$_2$ in the ideal 1T-structure.** The t$_{2g}$-p orbital coupling manifests itself in the overlap of their density of states (DOS). The local reference frame in the octahedral is used for projecting DOS. The DOS shown in the figure are t$_{2g}$ = d$_{xy}$ + d$_{yz}$ + d$_{zx}$, e$_g$ = d$_{x2-y2}$ + d$_{z2}$, and p = p$_x$ + p$_y$ + p$_z$. The vertical dashed lines show the position of Fermi-level. The energy is aligned to the vacuum level.

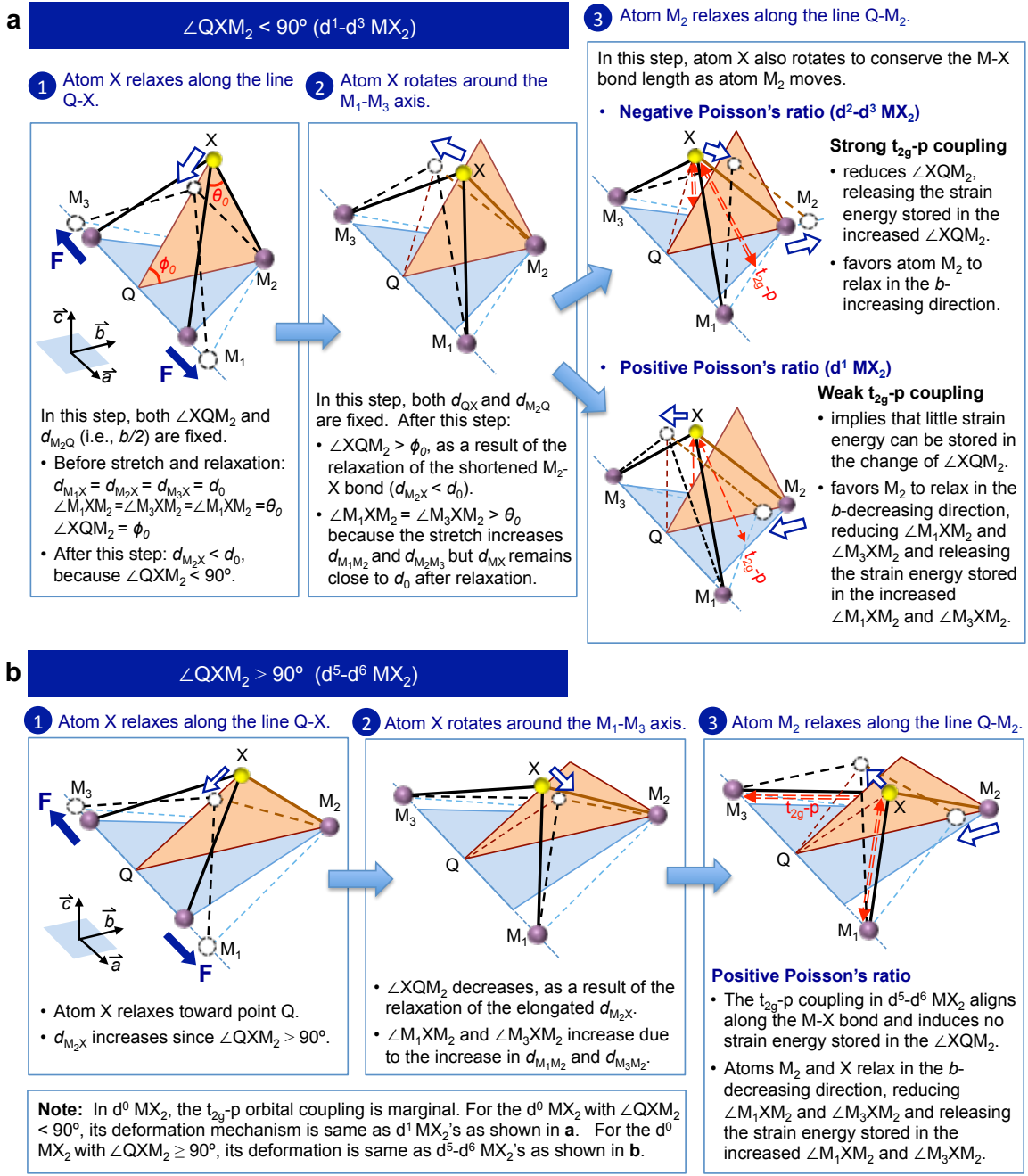

**Figure 4 | Deformation Mechanism.** The solid and dashed M-X bonds indicate, respectively, the initial and final configurations at each relaxation step. The force is applied along the lattice-*a* direction. The red dashed arrows indicate the direction of the $t_{2g}$-p orbital interaction. The hollow blue arrows show the resulting movement of the X and $M_2$ atoms within the Q-X-$M_2$ plane.

**Supplementary Table 1:** The Poisson's ratios and lattice constants of groups 6-7 $MX_2$ compounds calculated from SCAN and HSE06. Available experimental lattice constants are shown in parentheses.

| $M^{4+}$ \ $X_2$ | In-plane Poisson's ratio along $a$: SCAN/HSE06 | | | Lattice Constant $a$ (Å): SCAN/HSE06 (Experiment) | | |
|---|---|---|---|---|---|---|
| | $-S_2$ | $-Se_2$ | $-Te_2$ | $-S_2$ | $-Se_2$ | $-Te_2$ |
| Mo | -0.07/**0.01** | -0.12/-0.00 | -0.09/-0.01 | **3.1998/3.1657** (3.201)[1] | **3.2685/3.2529** (3.270)[1] | 3.4970/3.4626 (3.469)[2] |
| W | -0.04/**0.05** | -0.19/-0.05 | -0.20/-0.06 | **3.1908/3.1843** (3.222)[1] | 3.2574/3.2645 | **3.4970/3.4759** (3.496)[3] |
| Tc | -0.10/**0.03** | -0.04/**0.01** | -0.37/-0.02 | 3.0692/3.0255 | 3.1543/3.1386 | 3.4149/3.3486 |
| Re | -0.13/-0.02 | -0.03-0.02 | -0.22/-0.09 | 3.0750/3.0590 | 3.1311/3.1381 | 3.3834/3.3551 |

**Supplementary Table 2.** SCAN-calculated lattice constants of 42 1T-$MX_2$ compounds. Available experimental values are shown in parentheses. Note the experimental lattice constants listed for $d^5$-$d^6$ $MTe_2$ compounds are extracted from the distorted 1T structure. The small difference in lattice constant between the undistorted and distorted phases is expected.

| $M^{4+}$ | | $X_2$ | Lattice Constant $a$ (Å) | | |
|---|---|---|---|---|---|
| | | | $-S_2$ | $-Se_2$ | $-Te_2$ |
| $d^0$ | | Ti | 3.4055 (3.405)[4] | 3.5165 (3.535)[4] | 3.7648 (3.766)[4] |
| | | Zr | 3.6833 (3.662)[4] | 3.7815 (3.770)[4] | 4.0064 (3.950)[4] |
| | | Hf | 3.6153 (3.635)[4] | 3.7180 (3.748)[4] | 3.9606 (3.949)[5] |
| $d^1$ | | V | 3.2668 (3.29)[6] | 3.3260 (3.352)[4] | 3.6022 (3.595)[4] |
| | | Nb | 3.3870 | 3.4845 | 3.6738 (3.642)[7] |
| | | Ta | 3.3524 (3.346)[4] | 3.4602 (3.477)[4] | 3.6702 (3.651)[7] |
| $d^2$ | | Mo | 3.1998 (3.201)[1] | 3.2685 (3.270)[1] | 3.4970 (3.469)[2] |
| | | W | 3.1908 (3.222)[1] | 3.2574 | 3.4970 (3.496)[3] |
| $d^3$ | | Tc | 3.0692 | 3.1543 | 3.4149 |
| | | Re | 3.0750 | 3.1311 | 3.3834 |
| $d^5$ | | Co | Non-Layered Structure | | 3.5983 (3.802)[8] |
| | | Rh | | | 3.7563 (3.92)[8] |
| | | Ir | | | 3.8431 (3.928)[8] |
| $d^6$ | | Ni | 3.3174 | 3.4712 | 3.7248 (3.854)[8] |
| | | Pd | 3.5408 | 3.6759 | 3.9162 (4.037)[8] |
| | | Pt | 3.5237 (3.543)[4] | 3.6662 (3.728)[4] | 3.9554 (4.026)[8] |

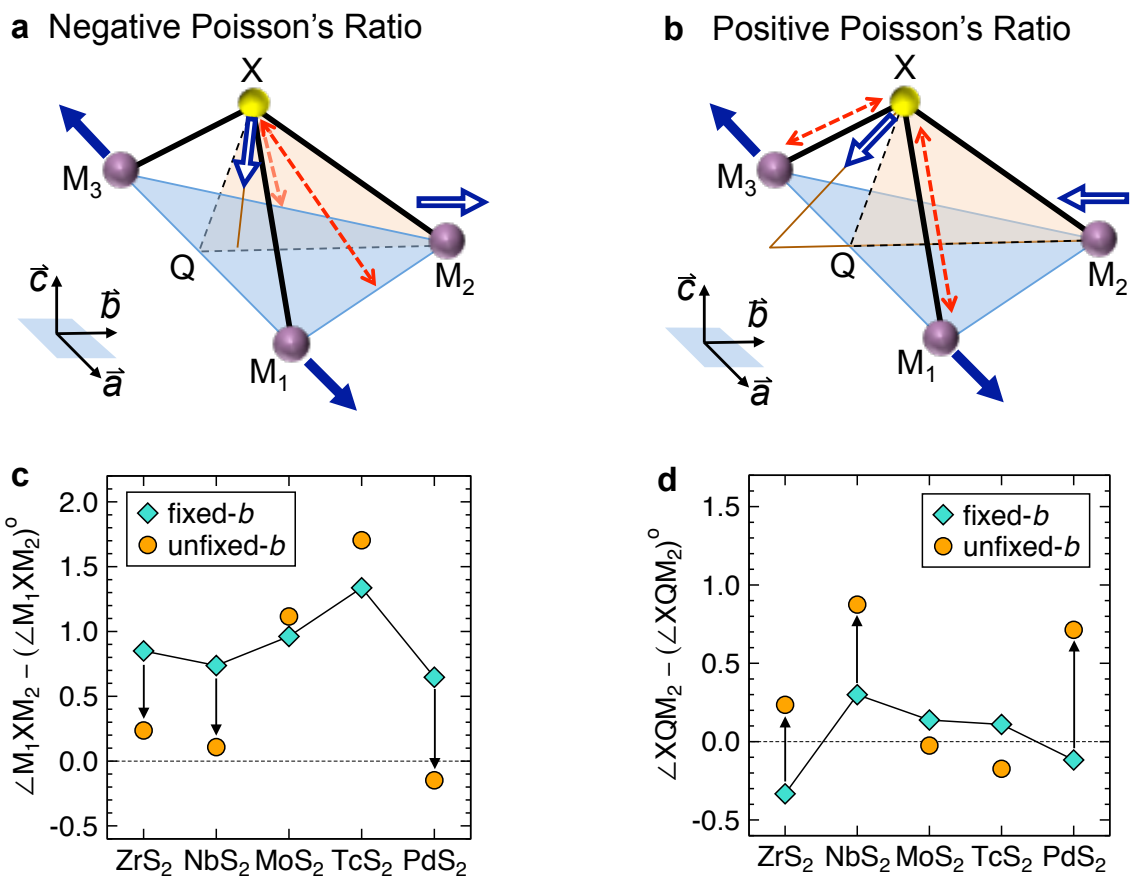

**Supplementary Figure 1 | Deformation Mechanism. a-b**, Motion of atoms in the auxetic and non-auxetic 1T-$MX_2$ compounds. The solid and open blue arrows indicate the applied force direction and resulting atom motion direction, respectively. The dashed red arrow indicates the direction of the $t_{2g}$-p orbital coupling force. **c,** Change in $\angle M_1XM_2$ after step-ii relaxation (i.e., the relaxation with the lattice $b$ fixed to its value in the relaxed strain-free structure). **d**, Change in X-$M_3$-$M_1$-$M_2$ dihedral angle $\angle QXM_2$ $\angle M_1XM_2$ after step-iii relaxation. The $(\angle QXM_2)^o$ and $(\angle M_1XM_2)^o$ are the reference angles taken in the relaxed structure of corresponding strain-free $MX_2$. The diamond and circle symbols represent the angles extracted from the structures relaxed under fixed lattice-$b$ and unfixed lattice-$b$, respectively, subjected to a 5% tension strain in the lattice-$a$ direction. The solid lines and arrows in c-d are guides to the eyes.